# Comparative Molecular Mechanics and Quantum Mechanics Study of Microhydration of Nucleic Acid Bases

J. Lino-Pérez, E. González-Jiménez, A. Deriabina, M. Velasco, V. Poltev

Facultad de Ciencias Físico Matemáticas, Benemérita Universidad Autónoma de Puebla, México, calle Río Verde y Av. San Claudio, C.U. Col. San Manuel, Puebla, 72570, Puebla, México

## Abstract

DNA is the most important biological molecule, and its hydration contributes essentially to the structure and functions of the double helix. We analyze the microhydration of the individual bases of nucleic acids and their methyl derivatives using methods of molecular mechanics (MM) with the Poltev-Malenkov (PM), AMBER and OPLS force fields, as well as *ab initio* Quantum Mechanics (QM) calculations at MP2/6-31G(d,p) level of theory. A comparison is made between the calculated interaction energies and the experimental enthalpies of microhydration of bases, obtained from mass spectrometry at low temperatures. Each local water-base interaction energy minimum obtained with MM corresponds to the minimum obtained with QM. General qualitative agreement was observed in the geometrical characteristics of the local minima obtained via the two groups of methods. MM minima correspond to slightly more coplanar structures than those obtained via QM methods, and the absolute MM energy values overestimate corresponding values obtained with QM. For Adenine and Thymine the QM local minima energy values are closer to those obtained by the PM potential (average of 0.72 kcal/mol) than by the AMBER force field (1.86 kcal/mol). The differences in energy between MM and QM results are more pronounced for Guanine and Cytosine, especially for minima with the water molecule forming H-bonds with two proton-acceptor centers of the base. Such minima are the deepest ones obtained via MM methods while QM calculations result in the global minima corresponding to water molecule H-bonded to one acceptor and one donor site of the base. Calculations for trimethylated bases with a water molecule corroborate the MM results. The energy profiles were obtained with some degrees of freedom of the water molecule being frozen. This data will contribute to the improvement of the molecular mechanics force fields.

# 1. Introduction

Water is one of the most abundant chemical compounds on the planet, and it constitutes a high percentage of the cell composition. To understand the role of interactions of biomolecules with water in relation to their functions, it is essential to have a detailed description of the energetic and structural aspects of interactions of the molecules involved. The first data on DNA fibers obtained by X-ray diffraction showed that DNA is highly hydrated and the interactions with water are responsible for its conformational changes [1]. The microhydration of nucleic acid (NA) bases, i.e. interactions of the bases with separate water molecules, plays an important role in structural stabilization of the double helix. QM calculations (of HF, DFT, and MP2) performed for hydrated complexes of DNA bases revealed that the geometric properties of such complexes are extremely sensitive to the interactions with one or few water molecules [2-4], the presence of just one water molecule is enough to completely change the structure of a complex of nucleic acid bases in the global minimum.

From the analysis of experimental results on hydration of oligomeric DNA duplexes, Schneider and his group [5,6] evaluated the distribution of water molecules around the components of the NA by considering it as a "construction of hydrated blocks". A modular scheme for the hydration was suggested. It determines the average sites of water molecules around the components of the NA, and can generate predictive patterns for the distribution of water molecules around the NA fragments. The studies of microhydration of the individual components of nucleic acids, obtained by both experimental and theoretical methods complete this scheme. Quantitative evaluation of the sites of hydration also contributes to the improvement of Molecular Mechanics force fields [7, 8].

Experimental spectroscopy studies have provided valuable data on the hydration of the components of the NA. The first studies of water clusters with nucleic bases using mass spectrometry in a primary ionization field were made by the group of Sukhodub, who determined the enthalpies of hydration of DNA bases and some of their derivatives [9]. Important mass spectroscopy experiments were performed by Kim et al. [10], where the threshold ionization energies for hydrated Adenine (A) and Thymine (T) bases were reported. Recent studies of UV photoionization in vacuum by a supersonic molecular beam using optical spectroscopy and comparison with theoretical results enabled the determination of the ionization energies of microhydrated DNA bases [11] and of tautomers of hydrated 9-methylguanine [12]. The studies of mononucleotide complexes with individual water molecules have been reported recently [13]. However, all these studies do not provide direct information about the structure and stabilization energy, and theoretical interpretation of the results is necessary. The experiments of Sukhodub et al. [9] represent the only exception, because they determine the gas-phase interaction energies of water-base complexes from the temperature dependences of the equilibrium constants of the association. However this method, in contrast to the theoretical results, does not specify the geometry of the complexes. So, despite the success of modern experimental methods, they still do not provide direct data for the detailed topology of the network of water-base hydrogen bonds (H-bonds). Thus, the computational methods of both molecular mechanics and quantum mechanics are the indispensable tools for the detailed study of the fine structure of hydration of nucleic acids.

The microhydration of the bases has been the subject of numerous theoretical studies by Monte Carlo [14-16] and Molecular Dynamics [17] techniques using different force fields. The hydration sites can be compared with quantum mechanics *ab initio* [12, 18-24] and DFT calculations for interactions of some water molecules with DNA bases and base pairs [4, 25-27]. The molecular mechanics calculations have demonstrated that the deepest minima of the interaction energy of a water molecule with nucleic bases correspond to the formation of a water bridge between two hydrophilic atoms of the base. Such a bridge can be formed in three different ways, namely between two H-bond acceptor centers of the base, between two donor centers, and one acceptor and one donor centers. The first of these scenarios was analyzed with *ab intio* quantum mechanical calculations more extensively (applying different basis sets), as this configuration in molecular mechanics corresponds to global minima for Guanine and Cytosine. We performed preliminary *ab initio* calculations using the bases with rigid geometry followed by the complete energy minimization in the space of all the degrees of freedom. We also performed the study of energy profiles, i.e. the dependences of the interaction energy on some water displacements around the base hydration sites fixing certain geometrical parameters. This information will contribute to future improvements in force fields. The comparison of theoretical MM and QM results with the experimental data, demonstrates that we need to reconsider the geometry of some minima positions for the force field parameters adjustments.

## 2. Method of calculation

The systems considered contain one of methylated nucleotide bases (1-methylpyrimidine or 9-methylpurine) and one water molecule. The starting geometries of the bases are the average structures obtained from X-ray experimental data in crystals, these geometries have been used in previous works [15, 28]. For simplicity we will name the above mentioned methylated bases simply as Adenine, Thymine, Cytosine, and Guanine. The calculations of the interaction energy were performed within the two schemes of molecular mechanics (MM) and quantum mechanics (QM). For MM calculations Poltev-Malenkov (PM) force field [7,8] was used along with the potentials implemented in the AMBER program [29,30], in both cases the interaction energy is calculated as the sum of pair interactions of all the atoms constituting the molecules. For the PM potential, each atom-atom interaction consists of a Coulomb term and of Lennard-Jones (or 6-12) one (Eq. 1). To describe the interaction between the atoms capable of forming hydrogen bonds, the 6-12 term is replaced by a 10-12 term (Eq. 2)

$$E(r_{ij}) = kq_iq_jr_{ij}^{-1} - A_{ij}\, r_{ij}^{-6} + B_{ij}\, r_{ij}^{-12} \quad (1)$$

$$E(r_{ij}) = kq_iqr_{ij}^{-1} - A_{ij}^{(10)}\, r_{ij}^{-10} + B_{ij}^{(10)}\, r_{ij}^{-12} \quad (2)$$

In these equations, k is a numerical constant, $q_i$, $q_j$ are the effective charges of atoms i and j respectively (calculated by semiempirical quantum chemistry methods and reproduced the experimental dipole moments of the molecules), $r_{ij}$ is the distance between the atoms. The coefficients $A_{ij}$, $B_{ij}$ and $A_{ij}^{(10)}$, $B_{ij}^{(10)}$ are adjustable parameters whose numerical values are the same as in previous articles [8, 28]. The AMBER potentials [29] take into account the intra-molecular terms (whose contributions are small) and do not contain the 10-12 terms.

Quantum mechanics calculations were performed using the GAUSSIAN 03W program [31], at MP2/6-31G(d, p). The interaction energies $E_{int}$ was evaluated considering the basis set superposition error correction using the counterpoised procedure of Boys-Bernardi implemented the GAUSSIAN package [32]. After energy minimization, additional single point calculations were performed with counterpoise option to evaluate the energy of the first molecule with the basic functions of the second one, and vice versa the energy of the second molecule with the functions of the first one. These terms are subtracted from the total energy and so the corrected energy $E_{BSSE}$ of the system is obtained.

$$E_{int} = E_{bsse} - E_{mol} - E_{water} \quad (3)$$

All the local minima were verified by the calculations of the matrices of second derivatives of energy (Hessian) which appeared to be positive. For some local minima of Guanine and Cytosine more extensive basis set (aug-cc-pvdz) was used in order to confirm the geometry. For each base, energy scans were performed with both methods (MM and QM) by changing the position of the water molecule around the hydrophilic centers. Some geometric parameters were varied gradually, with other ones being fixed. For example azimuthal scans were made, i.e. the angle θ (Fig. 3a.) was varied to change the position of the water molecule in the base plane around the base atom capable of forming a hydrogen bond. During these minimizations the distance r between two atoms of the two molecules and parameters □x, □y and □z which determine the rotations of the water hydrogen's around the water oxygen were varied (Fig. 3a.). The energy profiles obtained provide fine details of geometry changes which will contribute to the improvement of force fields.

## 3. Results and Discussion

### 3.1. Local minima of interaction energy between DNA bases and single water molecule

The extensive calculations of the water-base systems via MM and QM methods described in the previous section enable us to reveal all the local minima for these systems. The calculated interaction energies along with those of other authors are presented in Table 1. The hydrogen bond distances for these complexes are shown in Table 2.

The structures obtained with both methods are shown in Figure 1. The same number of local minima and rather close water oxygen positions were revealed in both MM and QM calculations, i.e. every MM minimum corresponds to QM minimum with the mutual water-base positions resembling those of QM minima. The PM potential functions favor the coplanar configurations of the complexes, i.e. the base ring and the 3 atoms of the water molecule located in nearly the same plane. The only exceptions are the configuration **3** for Guanine and the configuration **1** for Cytosine (Fig. 3). Qualitatively similar structures are obtained with the AMBER potentials and the potential of Jorgensen (OPLS), the latter values were calculated in reference [28] and coincide with the values obtained in [14] using the method of diffusion Monte Carlo. The results of our *ab initio* calculations revealed both non-coplanar and coplanar conformations, for example the structures corresponding to minimum **2** for Adenine and **3** for Cytosine are completely planar whereas for structures **2** and **3** for Thymine and **3** for Guanine, one of the hydrogens of the water molecule remains in the plane of the base while the other hydrogen deviates from the plane for approximately 30°.

Table 2 shows the inter-atomic distances of hydrogen bonds, the hydrogen bonds are shorter for MM than for QM. The inter-atomic distances N/O...$H_W$ and NH...$O_W$ for the potential PM fall in the range from 1.78 to 1.98 Å while for QM they vary from 1.94 to 2.24 Å. For AMBER potentials this region extends from 1.70 to 2.12 Å, i.e. it is larger than for PM but shorter than for the quantum-mechanical calculations deviating on average by 0.16 Å from the QM values.

Comparison of the interaction energies of the minima obtained with the MM method and those obtained with QM shows generally higher values for the former, this is true for both PM and AMBER potentials (not for all the OPLS results). This difference can be due to the MM potential adjustment to the hydration of the bases in aqueous solution [28, 8] where water molecules of the first shell are affected by the "bulk" water. The tendency for shortening the interatomic distances on including the other water molecules can be seen from comparison with other publications. This feature is reported in a DFT study [2] for complexes of Cytosine with 14 molecules of water, where for water position corresponding to minimum **4** the O-$H_W$...O2 distance is of 1,82 Å, while our QM calculations give the value of 1.91 Å. The same tendency took place in the Hartree-Fock study [19] for Guanine with 7 to 13 water molecules.

The values of the interaction energies in minima calculated with the method MM/PM are closer to QM ones for the Adenine and Thymine (the average differences being of 0.72 kcal/mol) than for Guanine and Cytosine (2.8 kcal/mol). The reason for these differences is due to the fact that QM calculations result in rather small interaction energies for H-bonding of water molecule to two proton acceptors of the bases. This situation will be discussed in the next section.

## 3.2. Interactions of water molecule with two H-bond acceptors of the bases

The most significant differences between MM and QM results refer to the minimum **1** of Guanine and **3** of Cytosine (Fig. 1) corresponding to the interaction of water molecule with two H-bond acceptors of the base. From calculations carried out with the potential PM, we found that: The first minimum for Guanine obtained via PM potentials is the most profound one, and it is only 0.2 kcal/mol deeper than the minimum **2**, which is the global one for AMBER force field (Table 1). The energy value for the minimum **1** obtained via QM calculations is less negative (-7.72 kcal/mol). The global QM minimum corresponds to the position **2** of (-10.81 kcal/mol). The interatomic distances for both QM and MM fall in the limits allowed by the geometric criterion for hydrogen bond formation [7] (Table 2), similar situation occurs for the minimum **3** for Cytosine, the water forms H-bonds via hydrogen atoms with two proton-acceptor centers of Cytosine; the QM distances N3...$H_w$ and O2…$H_W$ resemble corresponding distances for Guanine-water complex (2.13 and 2.31 Å, respectively). With MM methods similar energy values were obtained for both force fields (Table 1). Our global QM minimum 2 with the value of -10.24 kcal/mol resembles that obtained with more extensive basis set [4] and via DFT calculations [25,26] (the values of -9.97 and -9.1 kcal/mol respectively). Microhydration of Cytosine and its radical anion were investigated with the DFT-B3LYP method [26], and the deepest minimum for Cytosine was found at position **2**, but for the Cytosine anion it was located at position **3**, the bond lengths being shorter compared to our QM values, ($H_w$…O2 of 1.95 Å and $H_w$...N3 of 2.16 Å). The barrier between the minima **3** and **2** for cytosine-water complex is quite small; the minima **2** and **3** obtained via PM potentials have nearly the same energy (the difference of 0.1 kcal/mol). The same effect can be seen for different tautormers of Guanine via MM calculations [28]. The enol tautomer with the OH group oriented towards the N1 atom forms a complex with water molecule H-bonded to two acceptor atoms (N7 and O6) with the energy of -10.04 kcal/mol, and interatomic distances with O6 of 1.85 Å and with N7 of 1.97Å. This minimum was also revealed using *ab initio* RI-MP2 method with TZVPP basis set [18]. The deepest minimum for dCMP (B3LYP/6-31G* level of DFT) corresponds to the minimum **3** of Cytosine [33]. Thus, the MP2/6-31G(d,p) level of QM calculations results in the minima of poor stability when water molecule forms H bonds with two base acceptors, and the potential barrier is rather small, however, the MM potentials show them as ones of the lowest energy.

## 3.3. Microhydration of some methylated derivatives of nucleic acid bases

There are experimental mass spectrometry data [9] for some trimethylated bases which can be useful in comparison of the results of MM and QM methods on search for water-base interaction energy minima. The substitution of the hydrogen atom capable to form

H-bond by a methyl group excludes some local energy minima of water-base interactions, thus helping to refer experimental data to the definite water position. The methylation changes slightly the calculated values of water-base interaction energy minima for water positions not involved in H bonding with the hydrogen to be substituted. It was demonstrated for MM calculations earlier, and it is confirmed for QM calculations here.

The methylated bases considered in this section and compared with experimental results are: 1,4,4-trimetilcitosine ($m^{1,4,4}$Cyt), 2,2,9-trimetilguanine ($m^{2,2,9}$Gua), and 6,6,9-trimetiladenine ($M^{6,6,9}$Ade). The first one excludes the minima **1** and **2** for 1-methylcytosine, the second one excludes the minima **3** and **4** for 9-methylguanine, and the last one excludes the minima **1** and **3** for 9-methyladenine.

The calculation results obtained via MM and QM methods for trimethylated bases and the experimental enthalpies of water-base complex formation are listed in the Table 3. The values corresponding to global minima for 1-methylcytosine and 9-methylpurines (from the Table 1) are added for comparison.

The results demonstrate rather close experimental values of the enthalpies of complex formation with water molecule for $m^9$Gua and $m^{229}$Gua. The same is true for $m^1$Cyt and $m^{144}$Cyt (Table 3). Rather small differences between the values of monomethylated and trimethylated Gua and Cyt suggest the nearly same positions of the water molecule for complexes observed in experimental study and in global minima. The comparison of experimental data for $m^9$Ade and $m^{669}$Ade demonstrates less negative values for the trimethylated base, i.e. the substitution of amino group hydrogens by methyl groups changes the position of water molecule in the complex. Both MM and QM calculations suggest that the $m^{669}$Ade-water complex correspond to minimum **3** for $m^9$Ade, as the formation of other two minima for $m^9$Ade-water complexes are blocked by methyl groups.

The calculations for $m^{2,2,9}$Gua do not help to decide which minimum is more favorable, the minimum **1** (as predicted by PM potentials) or **2** (as predicted by QM and AMBER calculations). Both minima are possible for $m^{2,2,9}$Gua-water complex (Table 3), and the calculated values for these minima are close for those of $m^9$Gua (Table 1).

The calculations for $m^{144}$Cyt confirm the prediction of MM calculations (both PM and AMBER versions) on more favorable for $m^1$Cyt water position **3** (formation of two H bonds of water molecule with acceptors of the base) as compared to position **2** predicted from QM calculations. The position **2** is not possible for $m^{144}$Cyt-water complex, but experimental data demonstrate very close values of the enthalpy of hydration for $m^1$Cyt and $m^{144}$Cyt.

The calculations for trimethylated bases suggest the necessity of both improvement of MM force fields and more sophisticated QM calculations to reach more adequate description of water-base interactions.

## 3.4. Interaction energy dependences on displacement of water molecule from energy minima

Three types of scans via QM and MM methods described above have been performed for various displacements of water molecule from the positions corresponding to energy minima. The first type refers to azimuthal displacement of the water oxygen (other variables being free). The Figure 3 presents an example of such scans for Adenine-water complexes; Figure 3a demonstrates the regions of the $\varphi$ angle variations between N3, N1, and N7 acceptor sites. The dependences of interaction energy on the angle $\varphi$ demonstrate a similarity for MM and QM curves, and the energy values obtained with the two methods are rather close, especially for the regions near the minima **3** and **2**. The energy difference between two methods is largest in the region of the minimum **1** (1.15 kcal/mol, this value is different from that of Table 1, (0.39 kcal/mol) as we put constrains on some parameters). Other similarities of the two sets of curves refers to the distances between the atoms of two molecules; the dependencies of N...Ow and H...Ow distance on the $\varphi$ angle nearly coincide.

Similar azimuthal scans were obtained for other bases; for Thymine-water system maximum energy difference, 1.83 kcal/mol, corresponds to the trajectory from minimum **1** to minimum **2**. For the Guanine-water and Cytosine-water systems, there are more pronounced differences in energy, though the distances between the participating in H bonds atoms of the base and the water molecules are rather close for the two methods. It is noteworthy that for QM structures the distances of out-of-plane water hydrogen from base acceptor atom are nearly the same as for corresponding coplanar MM water-base complexes.

The second type of scans performed refer to moving a water molecule towards and away from the base starting from the minima positions (during the optimization $\phi x$, $\phi y$, and $\phi z$ parameters were varied, the angle $\varphi$ was fixed). When we make a radial scan such that a water molecule approaching the methyl group of the base to the distances between the oxygen and carbon shorter than 3.15 Å, the structures obtained with MM may be non-coplanar due to the repulsion of atoms. In this case the energy dependence as a function of **r** for the two methods demonstrate the same pattern.

The third type of scans was performed by the displacements of water molecules out of the plane of the bases, the angle $\theta z$ being varied from 0° to 90°. In this case the energies have the same tendency to decrease when the water moves away from the base plane (to $\theta z = 90$ °), at the end of the scan path there can arise a marked difference (up to 3 kcal/mol for scans near amino or methyl groups).

Some MM minima refer to both water hydrogen's in the base plane while corresponding QM minima refer to displacement of the water hydrogen not forming H-bond by 30°-45° out of the plane (e.g. in Thymine and Guanine 2nd minima). We performed MM and QM energy scans as functions of the angle of rotation about O-H water bond (H being H-bonded to the base and the bond being in the base plane). The energy differences between QM and MM water positions fall in 0.2 kcal/mol region, thus being not great, but may be significant for some cases. More profound QM calculations and MM parameter adjustment are required for more exact water-base system description in this respect.

## 4. Conclusion

This paper concerns the evaluation of the interactions of nucleic acid bases with single water molecule. The calculations for such simple systems can be performed via the methods of various complexities, from simple atom-atom MM computations of the rigid molecules to correlated *ab initio* QM computations using extended basis sets. The comparison of the results obtained via various methods demonstrates both some common features and some differences in quantitative geometry and energy characteristics. The simulation of biomolecular systems in surrounding water is possible via MM methods only. Thus, continuous improvement of MM force fields is required for adequate reproduction and prediction of important features of the systems containing nucleic acid fragments and hundreds of water molecules (and other biologically important molecules). Such improvement is not possible using experimental data only due to the insufficient amount of such data. The high level QM computations of the simple systems can help to fill this gap.

The comparison of the results of systematic QM MP2/6-31G(d,p) level computations with different MM methods is the first step on the pathway of MM force field refinement. Our MM computations using PM and AMBER force fields have demonstrated that each local MM energy minimum can be referred to QM one. The average energy difference between corresponding minima for Adenine and Thymine complexes with one water molecule is 0.72 and 1.86 kcal/mol for PM and AMBER force fields respectively. The differences for Guanine and Cytosine are more pronounced, especially for minima which correspond to the formation of two H bonds by water molecule with two acceptors of the bases. Such minima are global ones when calculated by MM methods while QM calculations results in global minima corresponding to the formation of one H-bonds with the base acceptor and another with base donor atom. The calculations for trimethylated bases and their comparison with experimental values of the enthalpy of monohydration supply us with evidences in favor to MM results. It became evident that additional and more extended computations via both more sophisticated QM methods and MM methods with changed force-field parameters are necessary for more exact description of base hydration. The comparison of QM and MM results for both energy minimum positions and energy dependences on selected variables should help to adjust the MM force field to the construction of detailed atom-level models of DNA fragments.

**Acknowledgements**

This work was partially supported by the VIEP-BUAP

# References

1. R.E. Franklin and R.G. Gosling, *Acta Crys.* **6**, (1953) 673.

2. O.V. Shishkin, L.Gorb and J. Leszczynski, *J.Phys.Chem. B* **104** (2000) 5357.

3. O.V. Shishkin, O. S. Sukhanov, L. Gorb and J. Leszczynski, *Phys.Chem. Chem.Phys.* **4** (2002) 5359.

4. T. van Mourik, V.I. Danilov, E. Gonzalez, A. Deriabina and V.I. Poltev, *Chemical Physics Letters* **445** (2007) 303.

5. B. Schneider, D.M. Cohen, L. Schleifer, A.R. Srinivasan, W.K. Olson and H. M. Berman, *Biophysical Journal* **65** (1993) 2291.

6. B. Schneider and H. M. Berman, *Biophysical Journal* **69** (1995) 2661.

7. V.I. Poltev, T.I. Grokhlina and G.G. Malenkov, *J.Biomol.Struc.Dyn*. **2, N2** (1984) 413.

8. E. González, F.I. Cedeño, A.V. Teplukhin, G.G. Malenkov and V.I. Poltev, *Rev.Mex.Fís.* **46, S2** (2000) 142.

9. L. F. Sukhodub, *Chem. Rev*. **87** (1987) 589.

10. S.K. Kim, W. Lee and D.R. Herschbach, *J. Phys. Chem*. **100** (1996) 7933.

11. L. Belau, K.R. Wilson, S.R. Leone and M. Ahmed, *J. Phys. Chem. A*, **111** (2007) 7562.

12. W. Chin. M. Mons, F. Piuzzi, B. Tardivel, I. Dimicoli, L. Gorb and J.Leszczynski, *J. Phys.Chem. A,* **108** (2004) 8237.

13. D. Liu, T. Wyttenbach and M.T. Bowers, *J.Am.Chem.Soc*. **128** (2006) 15155.

14**.** T. van Mourik, D.M. Benoit, S.L. Price and D.C. Clary, *Phys.Chem.Chem.Phys.* **2** (2000) 1281.

15. E. Gonzalez, A. Deriabina, A. Teplukhin, A. Hernández and V.I. Poltev, *Theo.Chem.Acc*., **110**, **N6** (2003) 460.

16. K. Coutinho, V. Ludwing and S. Canuto, *Phys. Rev. E*, 69, (2004) 61902.

17. M. Kabelác and P. Hobza, *Phys.Chem.Chem.Phys*., **9** (2007) 903.

18. M. Hanus, F. Ryjacek, M. Kabelac, T. Kubar, T.V. Bogdan, S.A. Trygubenko, P. Hobza *J.Am.Chem.Soc*. **125** (2003) 7678.

19. M.K. Shukla and J. Leszczynski *J.Phys.Chem. B*, **112** (2008) 5139.

20. B. Crews, A. Abo-Riziq, L. Grace, M. Callahan, M. Kabelac, P. Hobza and M.S. de Vries, *Phys. Chem. Chem. Phys*, **7** (2005) 3015.

21. M. Kabelac, L. Zendlova, D. Reha and P. Hobza, *J. Phys.Chem. B*. **109** (2005) 12206.

22. M. Hanus, M. Kabelac, J. Rejnek, F. Ryjacek and P. Hobza *J. Phys. Chem. B,* **108**, (2004) 2087.

23. S.A. Trygubenko, T.V. Bogdan, M. Rueda, M. Orozco, F.J. Luque, J. Sponer, P. Slavicek and P. Hobza, *Phys. Chem. Chem. Phys*. **4**, (2002) 4192.

24. J. Rejnek, M. Hanus, M. Kabelac, F. Ryjacek and P. Hobza, *Phys. Chem. Chem. Phys*. **7**, (2005) 2006.

25. S. Kim, S.E. Wheeler and H.F. Schaefer III, *J.Chem.Phys*. **124**, (2006) 204310.

26. S. Kim and H.F. Schaefer III, *J. Chem.Phys*., **126** (2007) 064301.

27. D.M. Close, C.E. Crespo-Hernández, L. Gorb and J. Leszczynski, *J.Phys.Chem. A*. **110**, (2006) 7485.

28. V.I. Poltev, G.G. Malenkov, E. Gonzalez, A.V. Teplukhin, R. Rein, M. Shibata and J.H. Miller, *J. Biomol. Struct. Dyn.* **13**, N4, (1996) 717.

29. W.D. Cornell, P. Cieplak, C.I. Gould, K.M. Jr. Merz, D.M. Ferguson, D.C. Spellmeyer, T. Fox, J.W. Caldwell and P.A. Kollman, *J. Am. Chem. Soc*. **117** (1995) 5179.

30. D.A. Case, T.E. III Cheatham, T. Darden, H. Gohlke, R. Luo, K.M. Jr. Merz, A. Onufriev, C. Simmerling, B. Wang and R. Woods. *J. Comput. Chem*. **26** (2005) 1668.

31. M. J. Frisch and et al. Gaussian 03, Revision C.02, Gaussian, Inc., Wallingford CT (2004).

32. S.F. Boys and F. Bernardi, *Mol. Phys*. **19 (**1970) 553.

33. K.C. Hunter, L.R. Rutledge and S.D. Wetmore, *J. Phys. Chem. A* **109** (2005) 9554.

Legends for Figures

Fig.1. The positions of local energy minima for nucleic acid bases complexes with water molecule obtained using MM (PM potentials), left, and QM, right, methods.

Fig.2. QM interaction energy minima for three methylated bases.

Fig.3. The azimuthal scans for three regions (designated at top) around Adenine base.

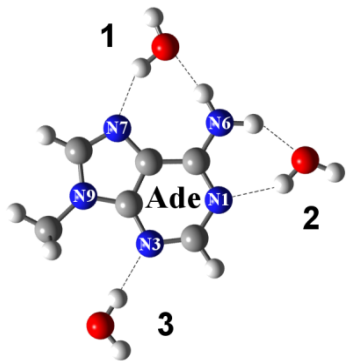
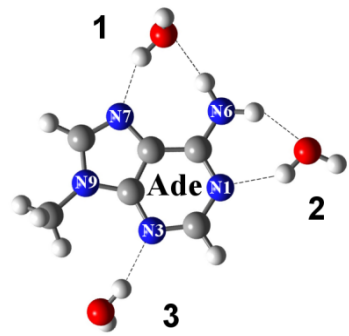

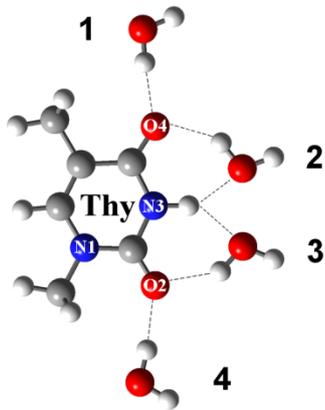
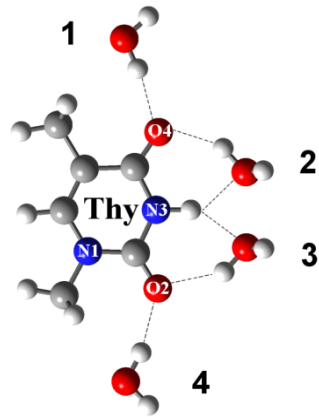

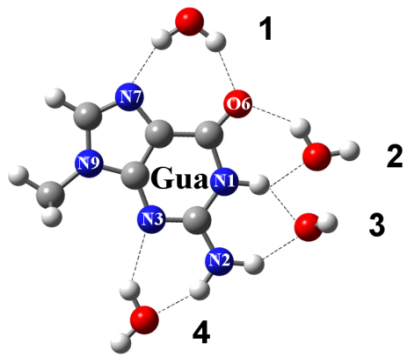
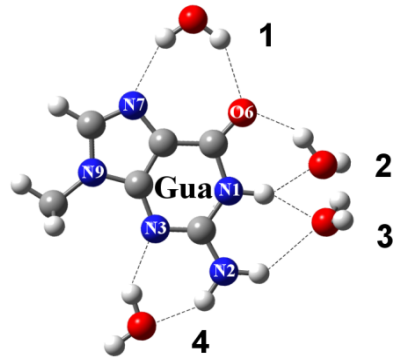

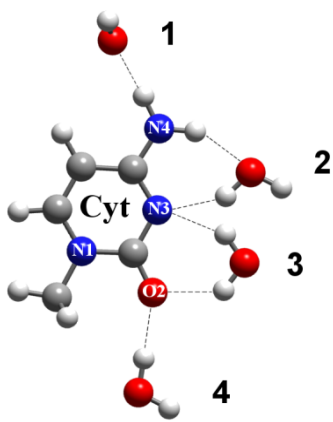
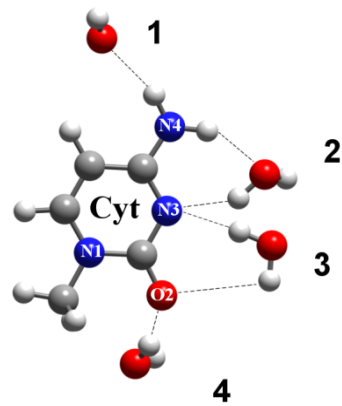

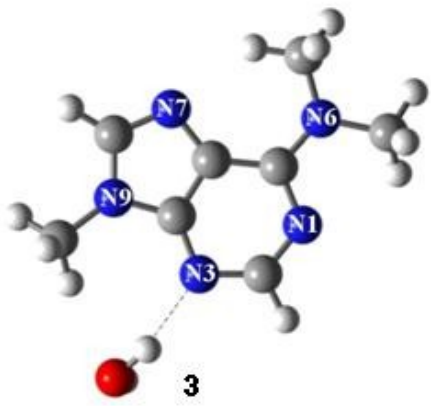 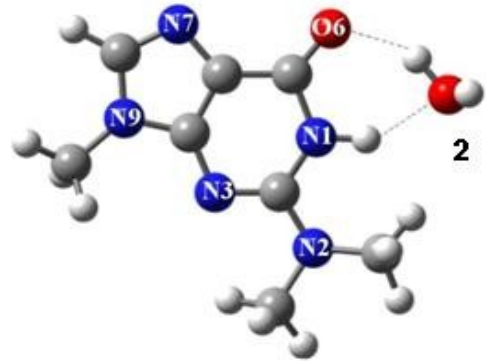 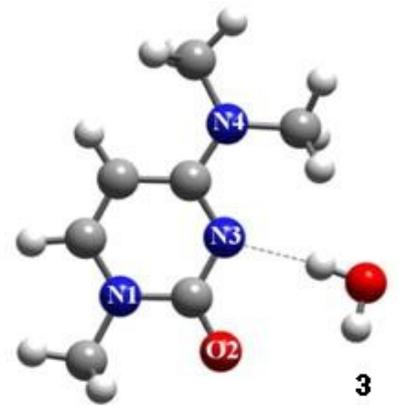

m$^{669}$Ade        m$^{229}$Gua        m$^{144}$Cyt

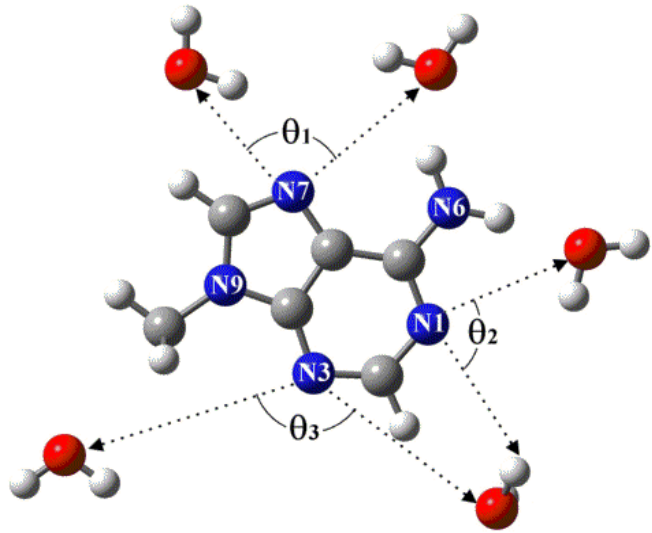

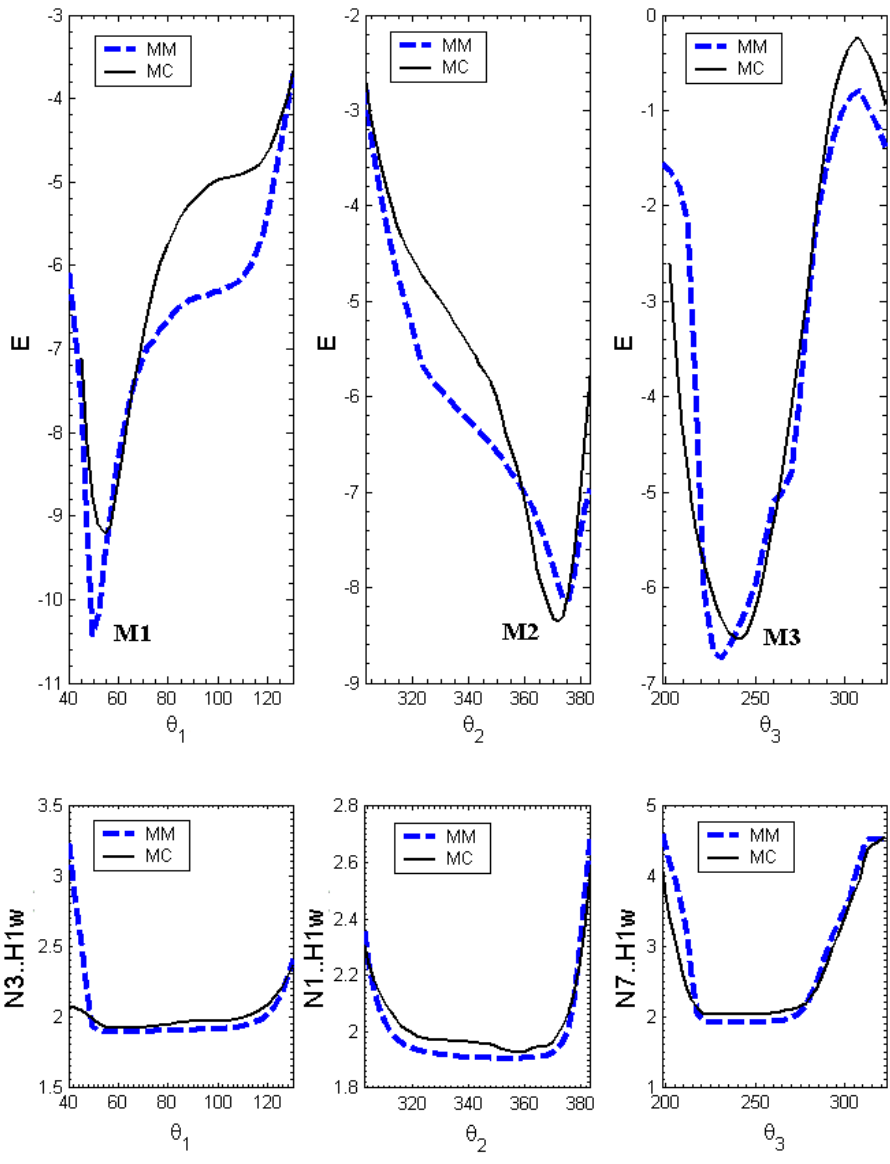

**Table 1**. Energies of water-base interactions (kcal/mol) in the local minima, calculated using different MM and QM methods.

| Minimum number | $E_{MP2}^{cal}$ | $E_{PM}$ | $E_{AMBER}$ | $E_{OPLS}$ | | $E_{MP2}^{*}$ | | $E_{DFT}$ | |
|---|---|---|---|---|---|---|---|---|---|
| 9-methyladenine ||||||||||
| 1 | -10.01 | -10.40 | -10.62 | -7.76 | | -9.3[b] | | | |
| 2 | -8.64 | -8.15 | -8.94 | -7.58 | | -8.7[b] | | | |
| 3 | -6.41 | -6.75 | -6.69 | -5.16 | | | | | |
| 1-metilthymine ||||||||||
| 1 | -4.92 | -6.96 | -8.9 | -6.40 | -6.97 | -5.9[b] | -5.65[a] | -4.6 | |
| 2 | -8.21 | -8.74 | -9.47 | -7.48 | -7.22 | -8.1[b] | -8.58[a] | -6.8 | |
| 3 | -7.81 | -8.05 | -10.47 | -6.46 | -6.73 | -8.2[b] | -8.35[a] | -6.5 | |
| 4 | -5.71 | -6.71 | -9.67 | -6.28 | | | | | |
| 9-methylguanine ||||||||||
| 1 | -7.72 | -11.98 | -9.99 | -9.37 | | | -7.31[c] | | |
| 2 | -10.81 | -11.78 | -12.20 | -11.11 | | -10.43[b] | -10.56[c] | | |
| 3 | -8.85 | -10.99 | -11.35 | -9.62 | | | -8.72[c] | | |
| 4 | -8.48 | -9.95 | -10.34 | -7.81 | | | -7.66[c] | | |
| 1-methylcytosine ||||||||||
| 1 | -5.88 | -7.56 | -6.38 | -6.24 | -6.41 | -5.24[a] | - | -4.5 | -4.47[d] |
| 2 | -10.24 | -10.81 | -7.82 | -9.92 | -9.85 | -9.97[a] | - | -9.1 | -8.26[d] |
| 3 | -7.39 | -10.91 | -11.69 | -8.75 | | | | | -5.06[d] |
| 4 | -6.32 | -8.35 | -5.46 | -7.33 | | | | | |

Structure numbering of the local minima corresponds to that of the Figure 1. $E_{MP2}^{cal}$ are the interaction energies calculated via MP2/6-31G(d,p) *ab initio* method. $E_{MP2}^{*}$ are the *ab initio* interaction energies calculated by other authors (MP2/6-31G(d,p) from ref 3([a]), RI-MP2 method from ref. 21-24 ([b]), MP2 dZ from ref. 14([c])). $E_{PM}$, $E_{AMBER}$ are the MM interaction energies calculated with PM and AMBER potentials respectively. $E_{OPLS}$ are MM energies obtained via OPLS potentials from ref.28 and 14 (second column). $E_{DFT}$ are the interaction energy obtained with DFT method by Kim et al [25, 26] and from ref. 33([d]).

**Table 2.** Hydrogen bond distances (Å) in the local minima of water-base interaction energy, calculated with Molecular Mechanics potentials PM y AMBER and with *ab initio* MP2/631G(d,p) method.

| Minimum number | Hydrophilic center | PM | AMBER | MP2/6-31G |
|---|---|---|---|---|
| | 9-methyladenine | | | |
| 1 | N7 | 1.94 2.83 | 1.82 2.79 | 1.92 2.84 |
| | N6-H62 | 1.80 2.73 | 1.88 2.86 | 1.94 2.92 |
| 2 | N6-H61 | 1.98 2.83 | 1.97 2.89 | 1.97 2.86 |
| | N1 | 1.88 2.77 | 1.83 2.79 | 2.01 2.90 |
| 3 | N3 | 1.91 2.87 | 1.84 2.81 | 1.99 2.93 |
| | 1-methyltimine | | | |
| 1 | O4 | 1.88 2.83 | 1.70 2.08 | 1.97 2.91 |
| 2 | O4 | 1.96 2.79 | 1.76 2.69 | 1.94 2.80 |
| | N3-H3 | 1.86 2.75 | 2.11 3.06 | 1.93 2.83 |
| 3 | O2 | 1.94 2.76 | 1.76 2.69 | 1.96 2.82 |
| | N3-H3 | 1.88 2.76 | 2.12 3.08 | 1.95 2.85 |
| 4 | O2 | 1.87 2.83 | 1.69 2.67 | 1.95 2.90 |
| | 9-methylguanine | | | |
| 1 | N7 | 1.91 2.80 | 2.01 2.94 | 2.16 3.04 |
| | O6 | 1.91 2.77 | 1.88 2.78 | 2.16 3.05 |
| 2 | N1-H1 | 1.84 2.76 | 2.00 2.69 | 1.89 2.81 |
| | O6 | 1.92 2.74 | 1.79 2.72 | 1.90 2.79 |
| 3 | N1-H1 | 1.91 2.76 | 2.03 2.96 | 2.43 3.25 |
| | N2-H21 | 1.88 2.76 | 2.07 2.97 | 1.94 2.92 |
| 4 | N2-H22 | 1.86 2.78 | 1.96 2.88 | 1.94 2.83 |
| | N3 | 1.98 2.80 | 1.85 2.80 | 1.98 2.83 |
| | 1-methylcitosine | | | |
| 1 | N4-H42 | 1.78 2.78 | 1.88 2.89 | 2.00 2.99 |
| 2 | N4-H41 | 1.91 2.81 | 1.98 2.90 | 1.96 2.88 |
| | N3 | 1.93 2.79 | 1.86 2.79 | 1.96 2.83 |
| 3 | N3 | 1.99 2.82 | 1.84 2.82 | 2.13 3.03 |
| | O2 | 1.96 2.68 | 2.55 3.00 | 2.31 3.02 |
| 4 | O2 | 1.86 2.82 | 1.67 2.66 | 1.92 2.85 |

The first value for each center corresponds to the N-H…O$_W$ o N/O$_{BASE}$…H$_W$ distance, and the second one to N/O$_{BASE}$…O$_W$ distance.

**Table 3**. Experimental enthalpies of formation and calculated water-base interaction energies (kcal/mol) of possible complexes for some methylated derivatives of the DNA bases.

| Structure | $\Delta H_{EXP}$ [9] | $E_{QM}$ | $E_{PM}$ | $E_{AMBER}$ | $E_{OLPS}$ [28] |
|---|---|---|---|---|---|
| $m^9$Ade | -10.6±1 | -10.01(**1**) | -10.40(**1**) | -10.62(**1**) | -7.76(**1**) |
| $m^{669}$Ade | -8.3±0.8 | -6.4(**3**) | -7.11(**3**) | -7.86(**3**) | -5.16(**3**) |
| $m^9$Gua |  | -10.81(**2**) | -11.98(**1**) | -12.20(**2**) | -11.11(**2**) |
| $m^{229}$Gua | -14±1 | -10.88(**2**) | -12.21(**2**) | -12.38(**2**) | -11.11(**2**) |
| $m^1$Cyt | -11.4±0.8 | -10.24(**2**) | -10.91(**3**) | -11.69(**3**) | -9.92(**2**) |
| $m^{144}$Cyt | -11.8±0.9 | -7.63(**3**) | -10.91(**3**) | -11.92(**3**) | -8.75(**3**) |

$\Delta H_{EXP}$, the experimentally obtained enthalpies (ref. 9). $E_{QM}$, the interaction energy calculated by *ab initio* MP2/6-31G(d,p) method. $E_{PM}$, $E_{AMBER}$, and $E_{OPLS}$ are designated as those values in the Table 1. The notations for the methylated bases are listed in the text. Numbers of minima according to Figure 1 are listed in parentheses.